\documentclass[a4paper,11pt]{article}
\pdfoutput=1
\usepackage{jcappub}

\usepackage[utf8]{inputenc}

\newcommand{\rR}{\rho_R}
\newcommand{\rbh}{\rho_\text{BH}}
\newcommand{\gs}{g_\star}
\newcommand{\gss}{g_{\star s}}
\newcommand{\Tbh}{T_\text{BH}}
\newcommand{\Tbhin}{T_\text{BH}^\text{in}}
\newcommand{\Mbh}{M_\text{BH}}
\newcommand{\tev}{t_\text{ev}}
\newcommand{\Tev}{T_\text{ev}}
\newcommand{\Tbev}{\bar T_\text{ev}}
\newcommand{\Min}{M_\text{in}}
\newcommand{\Tin}{T_\text{in}}
\newcommand{\tin}{t_\text{in}}
\newcommand{\mdm}{m_s}
\newcommand{\ndm}{n_\text{DM}}
\newcommand{\Ndm}{N_\text{DM}}

\newcommand{\Cn}{\mathcal{C}_n}
\newcommand{\lhs}{\lambda_{hs}}
\newcommand{\Tfo}{T_\text{fo}}
\newcommand{\Teq}{T_\text{eq}}
\newcommand{\Tc}{T_\text{c}}
\newcommand{\sv}{\langle\sigma v\rangle}

\title{Dark Matter in the Time of Primordial Black Holes}

\author[a]{Nicolás Bernal}
\author[b,\,c]{and Óscar Zapata}
\affiliation[a]{Centro de Investigaciones, Universidad Antonio Nariño\\
Carrera 3 este \# 47A-15, Bogotá, Colombia}
\affiliation[b]{Instituto de Física, Universidad de Antioquia\\
Calle 70 \# 52-21, apartado aéreo 1226, Medellín, Colombia.}
\affiliation[c]{Abdus Salam International Centre for Theoretical Physics\\Strada costiera 11, 34151, Trieste, Italy.}

\emailAdd{nicolas.bernal@uan.edu.co}
\emailAdd{oalberto.zapata@udea.edu.co}

\abstract{
Hawking evaporation of primordial black holes (PBH) with masses ranging from $\sim 10^{-1}$ to $\sim 10^9$~g can generate the whole observed dark matter (DM) relic density.
However, a second DM production mechanism, like freeze-out or freeze-in, could have also been active in the early universe.
Here we study the interplay of these mechanisms, focusing on the scenario where PBHs dominate the energy density of the universe, leading to a nonstandard cosmological era.
For concreteness, we use the singlet scalar DM model as an example for this analysis.
}

\begin{document}
\begin{flushright}
    PI/UAN-2020-683FT
\end{flushright}

\maketitle

\section{Introduction} 
There is compelling evidence of the existence of dark matter (DM), an unknown, non-baryonic matter component whose abundance in the universe exceeds the amount of ordinary matter roughly by a factor of five~\cite{Aghanim:2018eyx}.
Up to now, the only evidence about the existence of this dark component is via its gravitational interactions.
In this context, DM could have been gravitationally produced by Hawking evaporation of primordial black holes (PBH).
In fact, PBHs are formed from inhomogeneities in the early universe~\cite{Carr:1974nx}.
If their initial mass is below $\sim 10^9$~g, they disappear through Hawking evaporation~\cite{Hawking:1974sw} before Big Bang nucleosynthesis (BBN), and are poorly constrained~\cite{Carr:2009jm, Carr:2020gox, Papanikolaou:2020qtd} (some promising signatures are discussed in Refs.~\cite{Morrison:2018xla, Ito:2019wcb, Zagorac:2019ekv, Ejlli:2019bqj, Inomata:2020lmk, Hooper:2020evu, Domcke:2020yzq}). 
During their lifetime, PBHs radiate not only standard model (SM) particles but also hidden sector states, and in particular DM.
In this regard, PBH evaporation may have played a key role in the DM genesis~\cite{Green:1999yh, Khlopov:2004tn, Dai:2009hx, Fujita:2014hha, Allahverdi:2017sks, Lennon:2017tqq, Morrison:2018xla, Hooper:2019gtx, Chaudhuri:2020wjo, Masina:2020xhk, Baldes:2020nuv, Gondolo:2020uqv, Bernal:2020kse, Bernal:2020ili}.%
\footnote{See Refs.~\cite{Carr:2016drx, Carr:2020xqk, Green:2020jor} for recent reviews on PBHs as DM.}
Additionally, even if when forming PBHs were subdominant, as they scale like non-relativistic matter, their energy density will constitute an increasingly large fraction of the total energy density of the universe, naturally leading to a nonstandard cosmological era~\cite{Allahverdi:2020bys}.

However, despite the fact that the whole DM abundance can be created by PBHs, a second DM production mechanism could have also been active in the early universe.
In particular, if DM possesses a mass and coupling to the visible sector at the electroweak scale, it will be a WIMP generated via the the freeze-out paradigm~\cite{Steigman:1984ac, Steigman:2012nb, Arcadi:2017kky}.
Alternatively, if the couplings between the dark and visible sectors are very suppressed, so that DM never reaches chemical equilibrium with the SM, it could have been produced via the so-called freeze-in mechanism (FIMP)~\cite{McDonald:2001vt, Choi:2005vq, Hall:2009bx, Elahi:2014fsa, Bernal:2017kxu}.

In this paper, the interplay between DM production via PBH evaporation, and standard production mechanisms like the WIMP and the FIMP is analyzed.%
\footnote{The case of strongly interacting massive particles (SIMPs) was the subject of study in Ref.~\cite{Bernal:2020kse}.}
For concreteness, we focus on  the singlet scalar DM model as an example for this analysis; however, the conclusions could be generalized to other scenarios.
The singlet scalar DM model is therefore presented in section~\ref{sec:ssdm}, whereas in section~\ref{sec:PBH} we briefly review key aspects of PBH creation and evaporation.
Section~\ref{sec:DM} is devoted to the analysis of the different DM production mechanisms (PBH evaporation, FIMP, and WIMP) and their interplay, following a semi-analytical approach.
For the sake of completeness, in Appendix~\ref{sec:appendix} the relevant Boltzmann equations for the PBH energy density, the SM entropy density, and the DM number density are reported.
Finally, in section~\ref{sec:con} our conclusions are presented.

\section{Singlet Scalar Dark Matter} \label{sec:ssdm}
The singlet scalar model~\cite{Silveira:1985rk, McDonald:1993ex, Burgess:2000yq} is one of the minimal extensions of the SM that can provide a viable DM candidate. In addition to the SM framework, it only contains a real scalar $s$, singlet under the SM gauge group, but odd under a $\mathbb{Z}_2$ symmetry, which guarantees its stability. 
The most general renormalizable scalar potential is given by
\begin{equation}
    V = \mu_H^2\,|H|^2 + \lambda_H\,|H|^4 + \mu_s^2\,s^2 + \lambda_s\,s^4 + \lhs\,|H|^2\,s^2,
\end{equation}
where $H$ is the SM Higgs doublet.
The phenomenology of this model is completely determined by three parameters: the DM mass $\mdm$, the Higgs portal $\lhs$, and the DM quartic coupling $\lambda_s$.
Note that the role of the DM self-coupling $\lambda_s$ does not influence the DM abundance, if it is produced via the WIMP or the FIMP mechanisms.%
\footnote{We use \texttt{micrOMEGAs}~\cite{Belanger:2018ccd} to derive the WIMP and FIMP results in the radiation dominated era.}
However, it plays a major role in the case where DM is generated via the SIMP paradigm~\cite{Bernal:2015xba, Heikinheimo:2017ofk, Bernal:2020gzm}.

There has been a large amount of research on the singlet scalar DM model, most of them focused on the WIMP scenario, where the singlet $s$ has a sizeable mixing with the Higgs and undergoes a thermal freeze-out.
This scenario has been highly constrained by collider searches~\cite{Barger:2007im, Djouadi:2011aa, Djouadi:2012zc, Damgaard:2013kva, No:2013wsa, Robens:2015gla, Han:2016gyy}, DM direct detection~\cite{He:2009yd, Baek:2014jga, Feng:2014vea, Han:2015hda, Athron:2018hpc} and indirect detection~\cite{Yaguna:2008hd, Goudelis:2009zz, Profumo:2010kp, Cline:2013gha, Urbano:2014hda, Duerr:2015mva, Duerr:2015aka, Benito:2016kyp}.
In contrast, scenarios with a very suppressed Higgs portal are much less constrained, and could also lead to a vast phenomenology, such as the freeze-in mechanism~\cite{Yaguna:2011qn, Campbell:2015fra, Kang:2015aqa}.
Additionally, this model has also been studied in the framework of nonstandard cosmology~\cite{Bernal:2018ins, Hardy:2018bph, Bernal:2018kcw}.

\section{Primordial Black Holes: Formation and Evaporation} \label{sec:PBH}
Formation and evaporation of PBHs has been vastly discussed in the literature, see, for instance, Refs.~\cite{Carr:2009jm, Carr:2020gox, Masina:2020xhk, Gondolo:2020uqv}.
Here we briefly review the main aspects.
PBHs produced during the radiation dominated epoch, when the SM plasma has a temperature $T=\Tin$, have an initial mass $\Min$ of the order of the enclosed mass in the particle horizon
\begin{equation}
    \Min \equiv \Mbh(\Tin) = \frac{4\pi}{3}\, \gamma\, \frac{\rR(\Tin)}{H^3(\Tin)}\,,
\end{equation}
where $\gamma\simeq 0.2$, $\rR(T)\equiv\frac{\pi^2}{30}\,\gs(T)\,T^4$ is the SM radiation energy density with $\gs(T)$ the number of relativistic degrees of freedom contributing to $\rR$~\cite{Drees:2015exa}, and $H^2(T)=\frac{\rho(T)}{3M_P^2}$ is the squared Hubble expansion rate in terms of the total energy density $\rho(T)$, with $M_P$ the reduced Planck mass.

Once a PBH is formed, the evaporation process starts to radiate particles lighter than the BH temperature $\Tbh = M_P^2/\Mbh$~\cite{Hawking:1974sw}, leading to a mass loss rate~\cite{Page:1976df, Gondolo:2020uqv}
\begin{equation}\label{eq:dMdt}
    \frac{d\Mbh}{dt} = -\frac{\pi\gs(\Tbh)}{480}\frac{M_P^4}{\Mbh^2}\,.
\end{equation}
Therefore, the PBH mass evolution is given by
\begin{equation}
    \Mbh(t)=\Min\left(1-\frac{t-\tin}{\tau}\right)^{1/3},
\end{equation}
with $\tin$ being the time at PBH formation, and
\begin{equation}
    \tau\equiv\frac{160}{\pi\,\gs(\Tbh)}\frac{\Min^3}{M_P^4}\gg \tin
\end{equation}
the PBH lifetime.
Here the temperature dependence of $\gs$ during the BH evolution has been neglected.
If the universe remains dominated by SM radiation along the BH lifetime, it follows that SM plasma temperature when the BH has completely faded away is
\begin{equation}\label{eq:Tev}
    \Tev \equiv T(\tev)\simeq \left(\frac{9\,\gs(\Tbh)}{10240}\right)^\frac14 \left(\frac{M_P^5}{\Min^3}\right)^\frac12.
\end{equation}
However, if the PBH component dominates at some point the total energy density of the universe, the SM temperature just after the complete evaporation of PBHs is $\Tbev=\frac{2}{\sqrt{3}}\Tev$.  

The total number $N_j$ of the species $j$ of mass $m_j$ emitted during the PBH evaporation is given by
\begin{equation}\label{eq:N}
    N_j=\frac{15\,\zeta(3)}{\pi^4}\frac{g_j\,\Cn}{\gs(\Tbh)} \times
    \begin{cases}
        \left(\frac{\Min}{M_P}\right)^2\qquad\text{for}\quad m_j\leq\Tbhin\,,\\[8pt]
        \left(\frac{M_P}{m_j}\right)^2 \qquad\text{for}\quad m_j\geq\Tbhin\,,
    \end{cases}
\end{equation}
where $\Tbhin\equiv\Tbh(t=\tin)$ is the initial PBH temperature, and $\Cn = 1$ or $3/4$ for bosonic or fermionic species, respectively.
We note that the radiated particles are relativistic, with a mean energy $\langle E_j\rangle = 6\times \max(m_j,\,\Tbhin)$.

The initial PBHs abundance is characterized by the dimensionless parameter $\beta$ 
\begin{equation}
    \beta\equiv\frac{\rbh(\Tin)}{\rR(\Tin)}\,,
\end{equation}
which corresponds to the ratio of the initial PBH energy density to the SM energy density at the time of formation $T=\Tin$.
We note that $\beta$ steadily grows until evaporation, given the fact that BHs scale like non-relativistic matter ($\rbh \propto a^{-3}$, with $a$ being the scale factor), while $\rR \propto a^{-4}$.
A matter-dominated era (i.e., a PBH domination) can be avoided if $\rbh\ll\rR$ at all times, or equivalently if $\beta \ll \beta_c \equiv \Tev/\Tin$\,.

It has been recently pointed out that the production of gravitational waves (GW) induced by large-scale density perturbations underlain by PBHs could lead to a backreaction problem.
However, it could be avoided if the energy contained in GWs never overtakes the one of the background universe~\cite{Papanikolaou:2020qtd}:
\begin{equation}\label{eq:GW}
    \beta < 10^{-4}\left(\frac{10^9~\text{g}}{\Min}\right)^{1/4}.
\end{equation}

Finally, PBH evaporation produces all particles, and in particular extra radiation that can modify successful BBN predictions.
To avoid it, we require PBHs to fully evaporate before BBN time, i.e., $\Tev>T_\text{BBN}\simeq 4$~MeV~\cite{Sarkar:1995dd, Kawasaki:2000en, Hannestad:2004px, DeBernardis:2008zz, deSalas:2015glj}, which translates into an upper bound on the initial PBH mass.
On the opposite side, a lower bound on $\Min$ can be set once the upper bound on the inflationary scale is taken into account: $H_{I} \leq 2.5\times10^{-5}M_P$~\cite{Akrami:2018odb}.
Therefore, one has that
\begin{equation}\label{eq:PBHmass}
    0.1~\text{g}\lesssim \Min \lesssim 2\times 10^8~\text{g}\,.
\end{equation}

\section{Dark Matter Genesis} \label{sec:DM}
Even in the case where the dark and visible sectors are disconnected, DM can be produced via gravitational processes.
PBHs can radiate DM particles promptly or emit mediator states that later decay into DM.\footnote{We consider that the BH evaporation process does not cease at $\Tbh\sim M_P$, avoiding therefore the production of Planck mass relics~\cite{MacGibbon:1987my, Barrow:1992hq, Carr:1994ar, Dolgov:2000ht, Baumann:2007yr, Hooper:2019gtx, Dvali:2020wft}.
Mixed DM scenarios comprising PBHs and WIMPs are discussed in Refs.~\cite{Lacki:2010zf, Eroshenko:2016yve, Boucenna:2017ghj, Adamek:2019gns, Bertone:2019vsk, Carr:2020mqm, Gondolo:2020uqv}.}
Additionally, the $s$-channel exchange of gravitons gives an irreducible contribution that can be dominant for super heavy DM.
However, if there exist additional portals, DM can also be generated via other mechanisms like the WIMP or FIMP paradigms.
In the following, the interplay between these options will be studied in the framework of the singlet scalar DM model.

\subsection{PBH Evaporation}
The whole observed DM relic abundance could have been Hawking radiated by PBHs. 
The DM production can be analytically computed in two limiting regimes where PBHs dominate or not the energy density of the universe, and will be presented in the following.

The DM yield $Y_\text{DM}$ is defined as the ratio of the DM number density $\ndm$ over the SM entropy density $s(T)\equiv\frac{2\pi^2}{45}\gss(T)\,T^3$, where $\gss(T)$ is the number of relativistic degrees of freedom contributing to the SM entropy~\cite{Drees:2015exa}, at present.
For the DM production by Hawking evaporation of PBHs, $Y_\text{DM}$ can be estimated by~\cite{Masina:2020xhk, Gondolo:2020uqv, Bernal:2020kse}
\begin{equation}
    Y_\text{DM} \equiv \frac{\ndm(T_0)}{s(T_0)}
     \simeq \frac34\, \frac{\gs(\Tbh)}{\gss(\Tbh)}\, \Ndm \times
     \begin{cases}
        \beta\,\frac{\Tin}{\Min} \quad &\text{for RD, }\beta \leq \beta_c\,,\\[8pt]
        \frac{\Tbev}{\Min} \quad &\text{for MD, }\beta \geq \beta_c\,,
     \end{cases}
\end{equation}
with $T_0$ the SM temperature at present, and $\Ndm = \Ndm^\text{prompt} + 2\,N_h\times \text{Br}(h\to 2\,$DM$)$ is the sum of the prompt DM production and the secondary coming from the decay of radiated Higgs bosons.
We notice that the DM yield has a smooth transition between the two eras, when $\beta \to \beta_c$ with $\Tbev=\Tev$.

In the case where PBHs manage to dominate the total energy density, the production of SM radiation via PBH evaporation efficiently dilutes DM.
The dilution factor $D$ is defined by
\begin{equation}
    D = \frac{Y_\text{DM}^\text{RD}}{Y_\text{DM}^\text{MD}} \simeq \beta\,\frac{\Tin}{\Tbev} \geq 1\,.
\end{equation}
Large values of $\beta$ and small $\Tin$ maximize the dilution.
The constraints coming from BBN and GWs (Eqs.~\eqref{eq:PBHmass} and~\eqref{eq:GW}, respectively), imply the upper bound $D \lesssim 10^{10}$.

Finally, we note that the capture of DM by PBHs becomes important for very heavy DM and a large initial abundance of PBHs, $\beta\gtrsim10^{-4}$~\cite{Gondolo:2020uqv}.
In such a case, DM  capture rate can be comparable to its production rate, and therefore a suppression in the DM relic abundance is expected. 

\subsection{FIMP}
\begin{figure}
	\centering
	\includegraphics[scale=0.51]{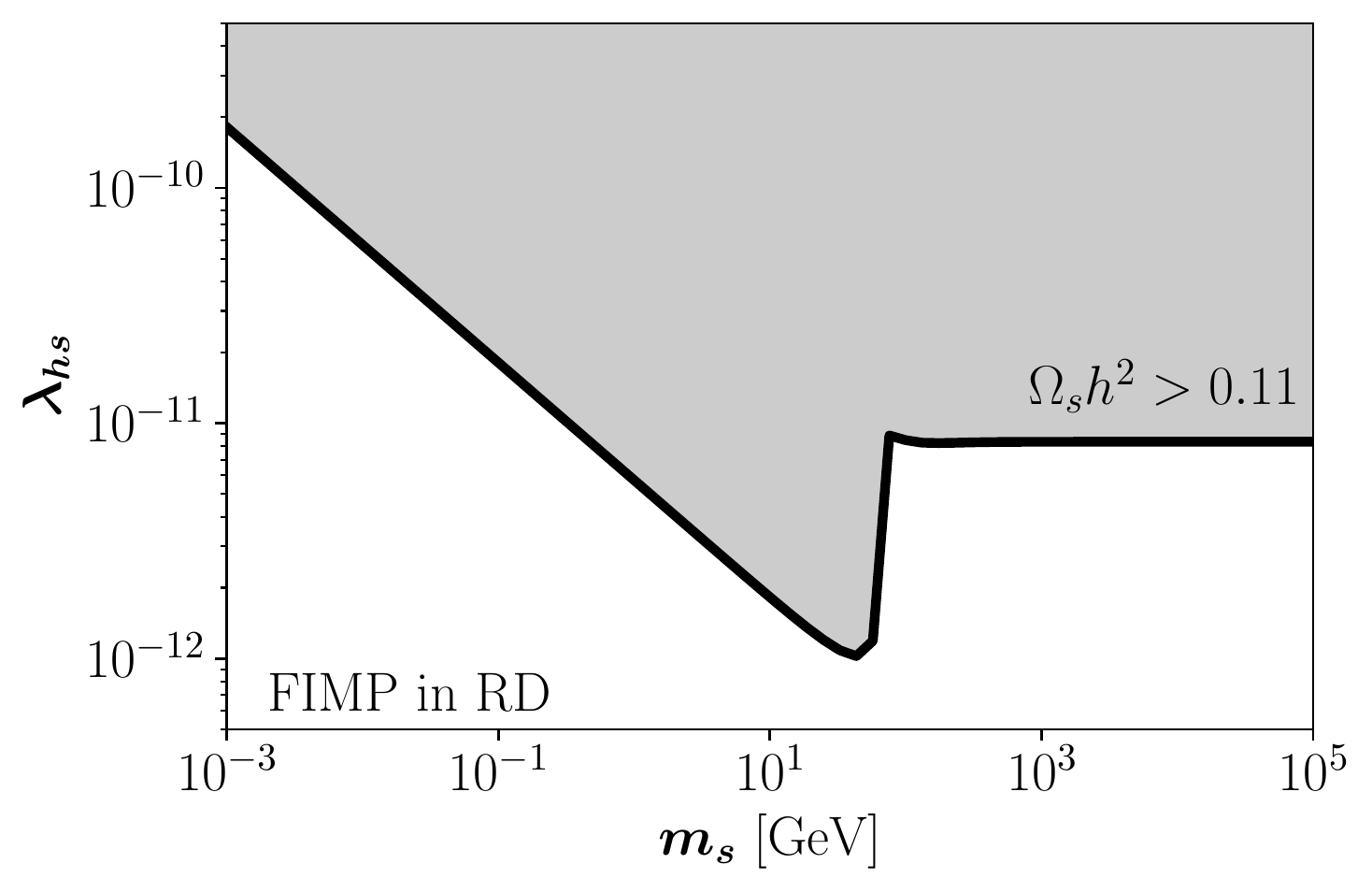}
	\caption{Portal coupling $\lhs$ required to reproduce the observed DM relic abundance as a function of the DM mass $\mdm$, in the freeze-in scenario, assuming a standard cosmology (black line).
	Higher couplings overclose the universe and are therefore excluded (gray area).}
	\label{fig:fimpRD}
\end{figure}

Alternatively, the whole DM abundance can also be produced via the FIMP paradigm, from 2-to-2 annihilations of SM particles, and decays of Higgs bosons~\cite{Yaguna:2011qn}.
The dominant production channels correspond to $W^+W^-\to ss$  for $\mdm \gg m_h/2$, and to $h\to ss$ for $\mdm \ll m_h/2$, where $m_h$ denotes the Higgs boson mass.
As expected in the FIMP scenario, the final DM yield is proportional to the production rate, and hence to $\lhs^2$.
Additionally, even if DM is continuously generated in the early universe, the bulk of its production happens at $T_\text{fi} \simeq$ max$(\mdm,\,m_h/2)$.
The required values for the coupling $\lhs$ as a function of the DM mass, in the standard case of a universe dominated by SM radiation, are of the order of $\lhs \sim {\mathcal O}\left(10^{-11}\right)$, and are shown in Fig.~\ref{fig:fimpRD} with a thick black line.
The gray-shaded area corresponding to larger couplings represents DM overproduction and is therefore excluded.

\begin{figure}
	\centering
	\includegraphics[scale=0.51]{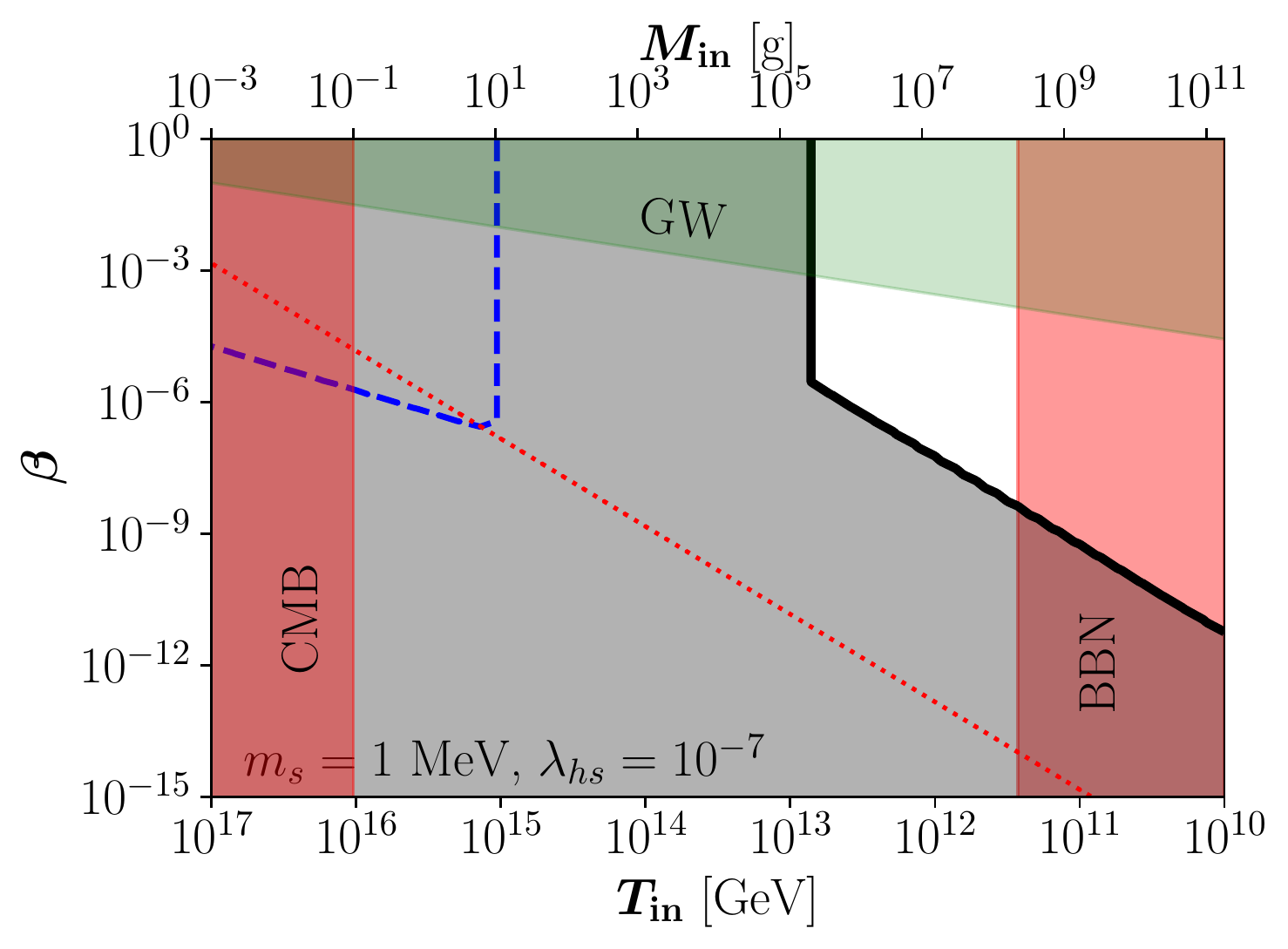}
	\includegraphics[scale=0.51]{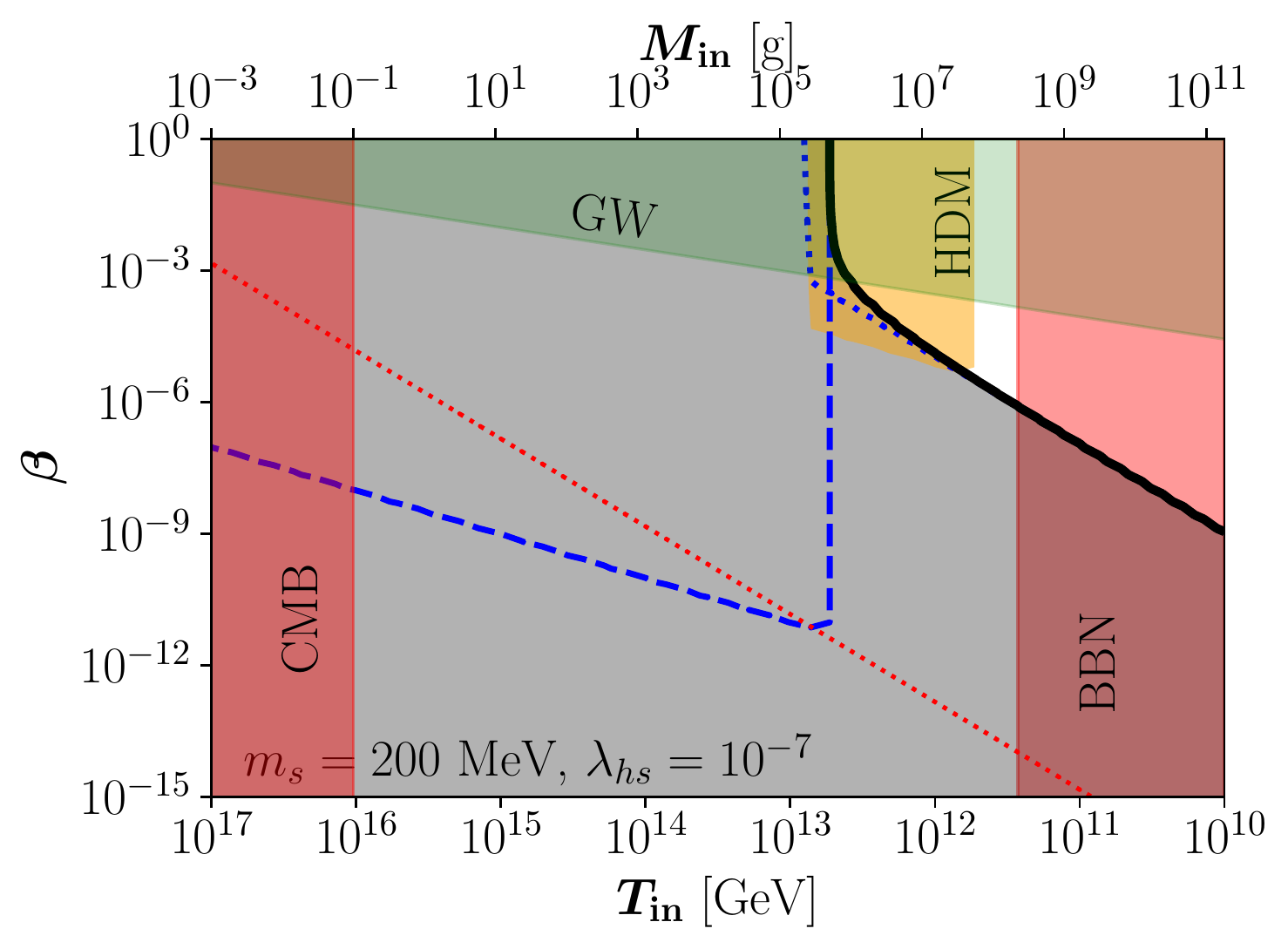}
	\includegraphics[scale=0.51]{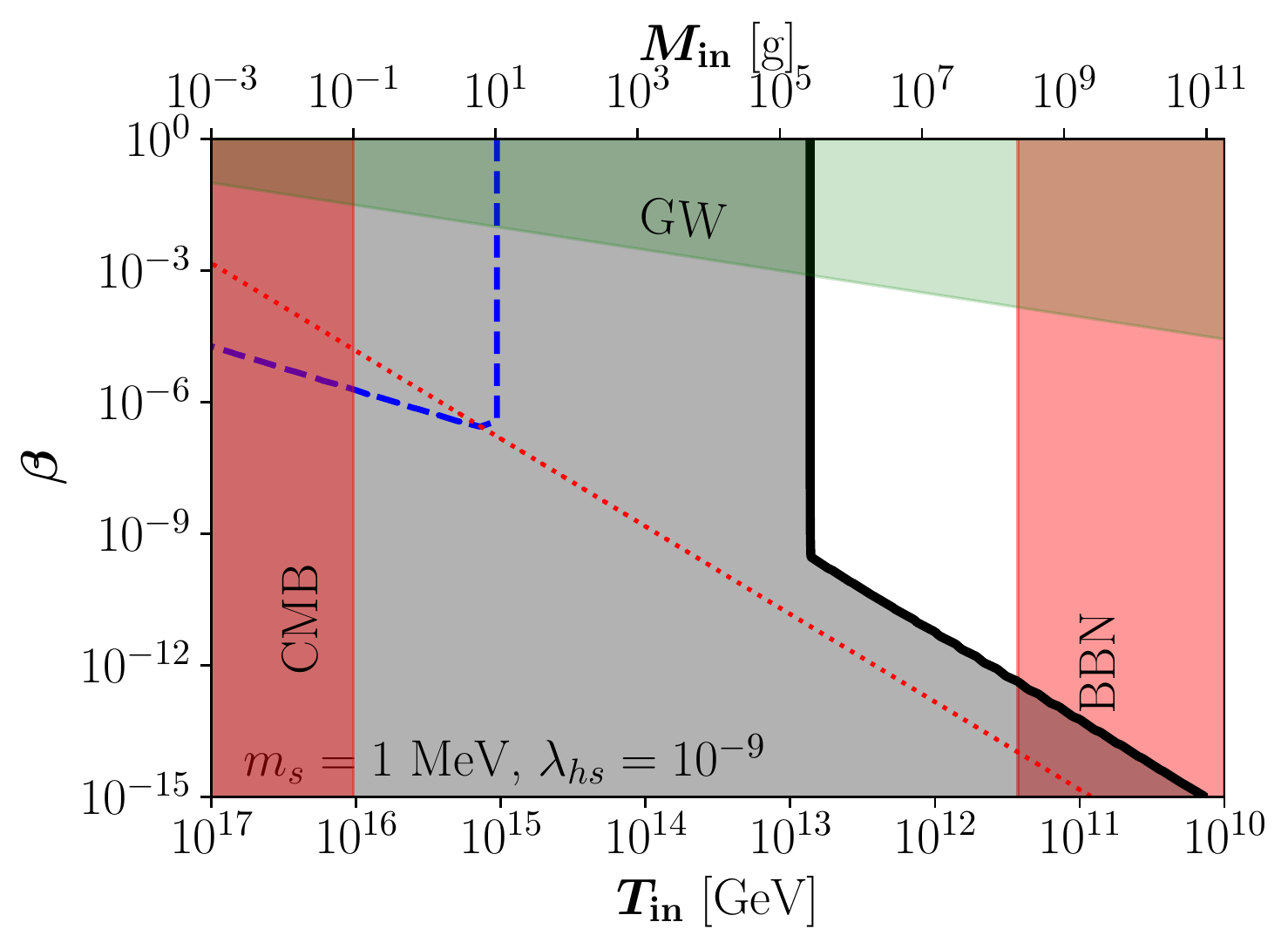}
	\includegraphics[scale=0.51]{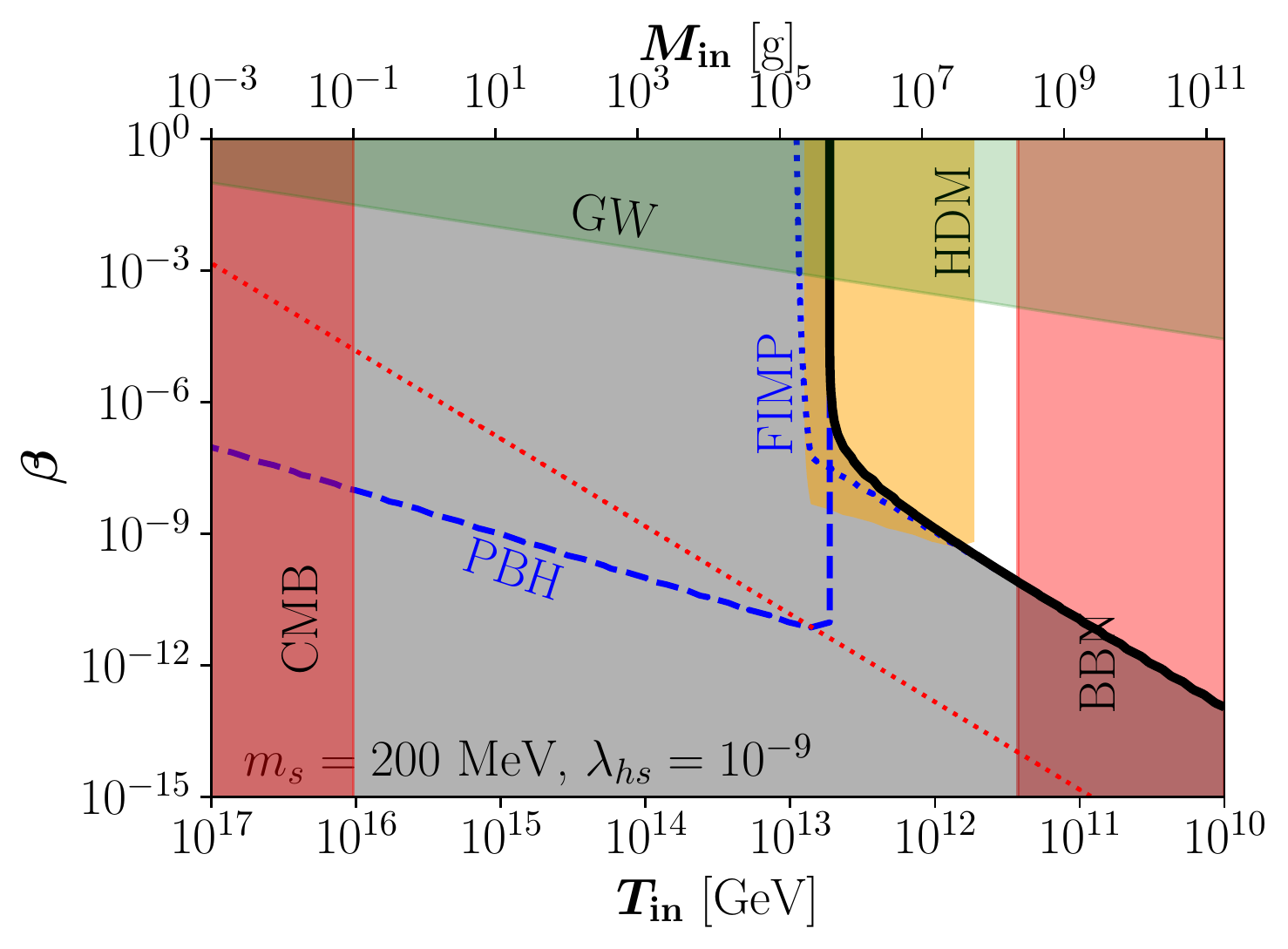}
	\caption{DM relic abundance due to the direct evaporation of PBHs (dashed blue lines) and the FIMP mechanism (dotted blue lines), for $\mdm = 1$~MeV (left panels) and 200~MeV (right panels), and $\lhs = 10^{-7}$ (upper panels) and $10^{-9}$ (lower panels).
	The shaded areas are excluded by different observables described in the text}
	\label{fig:fimp}
\end{figure}

In the case where PBHs never dominate the total energy density ($\beta < \beta_c$), the final DM abundance is simply the sum of the production by PBHs and the FIMP mechanism.
However, in the case opposite case where there is a PBH dominated era ($\beta > \beta_c$), entropy injection dilutes the produced DM, and therefore larger Higgs portal couplings are typically required.
However, for the dilution to have an impact on the final DM abundance, evaporation has to happen after the freeze-in, i.e., $\Tev < T_\text{fi}$.
For light DM ($\mdm < m_h/2$), that implies that dilution is only effective if $\Min \gtrsim 2\times 10^5$~g.
Figure~\ref{fig:fimp} shows the contributions to the DM relic abundance due to the direct evaporation of PBHs (dashed blue lines) and the FIMP mechanism (dotted blue lines), for $\mdm = 1$~MeV (left panels) and 200~MeV (right panels), and $\lhs = 10^{-7}$ (upper panels) and $10^{-9}$ (lower panels).
The values for the couplings were chosen so that they are higher than the required values for a radiation-dominated universe, i.e., $\lhs \simeq 2\times 10^{-10}$ for $\mdm =1$~MeV and $\lhs \simeq 10^{-11}$ for $\mdm =200$~MeV.
Additionally, much higher couplings can not be explored because thermalization with the SM is reached, and therefore the freeze-in picture is not consistent~\cite{Bernal:2018kcw}.
In this figure, the thick black lines correspond to the total contribution.
We notice that as the portal couplings are in this case very suppressed, the production via the decay of Higgs bosons emitted by PBHs is subdominant and does not appear in the figure.
We also note that for $\mdm \lesssim 100$~MeV, DM is mainly produced via the FIMP mechanism.
Additionally, in this figure the shaded regions represent parameter spaces constrained by different observables: CMB and BBN (both in red), GWs (green), and DM overabundance (gray).
The red dotted lines correspond to $\beta = \beta_c$ and therefore to the transition between a universe always dominated by SM radiation (below), or eventually dominated by PBH energy density (above).\\
Finally, this scenario corresponds to mixed DM, where a cold component is produced by the FIMP mechanism, and a (potentially) hot component is radiated by PBHs.
Data from CMB and BAO, and the number of dwarf satellites in the Milky Way, allow to measure the suppression of the matter power spectrum at the smallest scales due to the free-streaming of the noncold DM component.
The fraction of the noncold component with respect to the total DM can be bounded as a function of the DM mass~\cite{Boyarsky:2008xj, Kamada:2016vsc, Diamanti:2017xfo, Gariazzo:2017pzb}.
Here however, we take a conservative approach, and impose an upper bound of $10\%$, overlaid with orange in Fig.~\ref{fig:fimp}.
This constraint only applies to DM mass $\mdm \gtrsim 100$~MeV, when PBHs could produce a sizeable portion of the DM.
Additionally, for masses 400~MeV~$\lesssim \mdm \lesssim$~4~GeV, the DM produced in a PBH dominated era is too hot, and therefore in conflict with structure formation.

In the case where the DM mass spans in the range 4~GeV~$\lesssim \mdm \lesssim 10^9$~GeV, the BBN bound makes not viable the DM production in an early PBH dominated era.
For these masses, however, DM can be produced via the FIMP paradigm within the standard cosmology.

\begin{figure}
	\centering
	\includegraphics[scale=0.51]{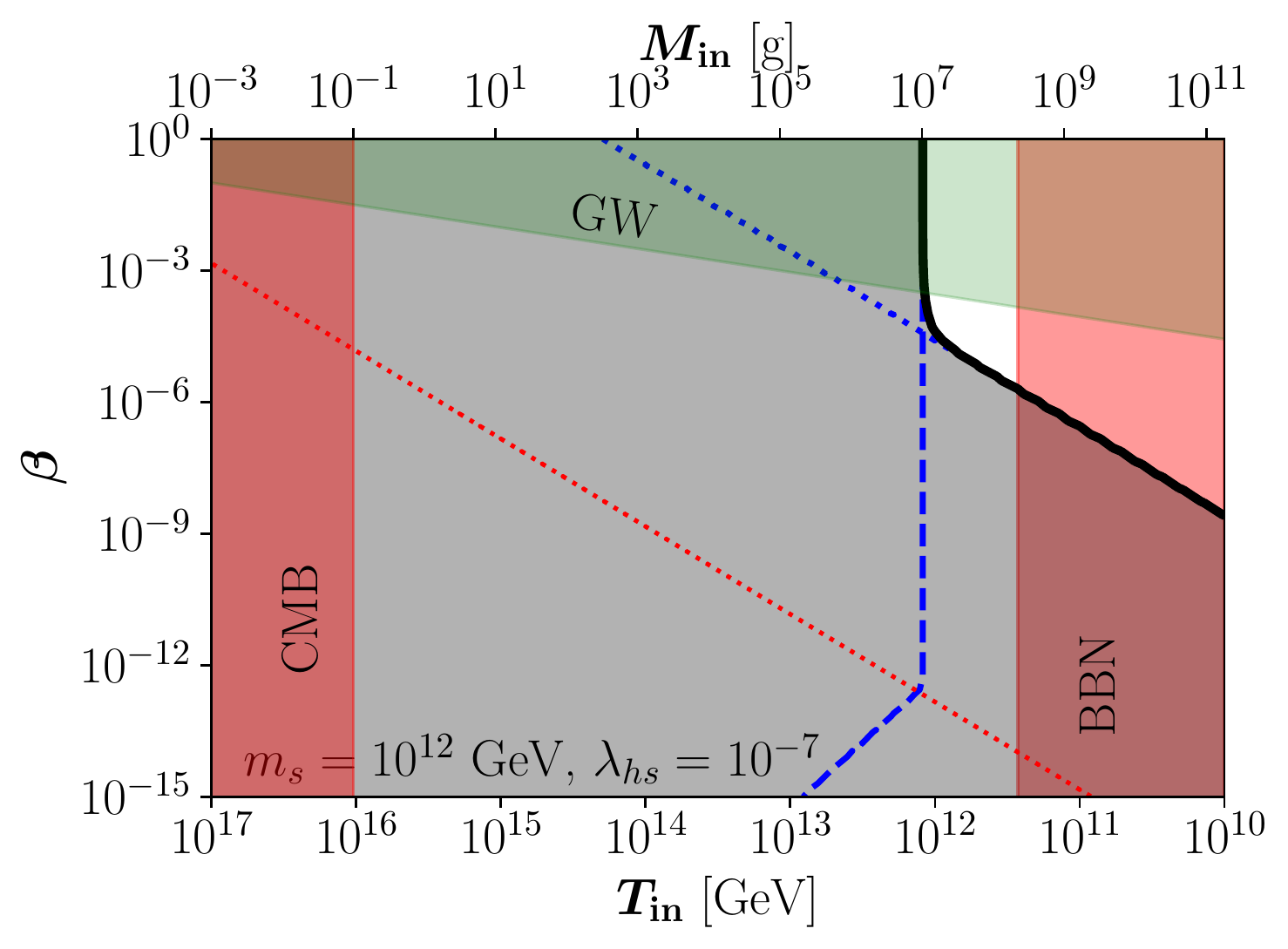}
	\includegraphics[scale=0.51]{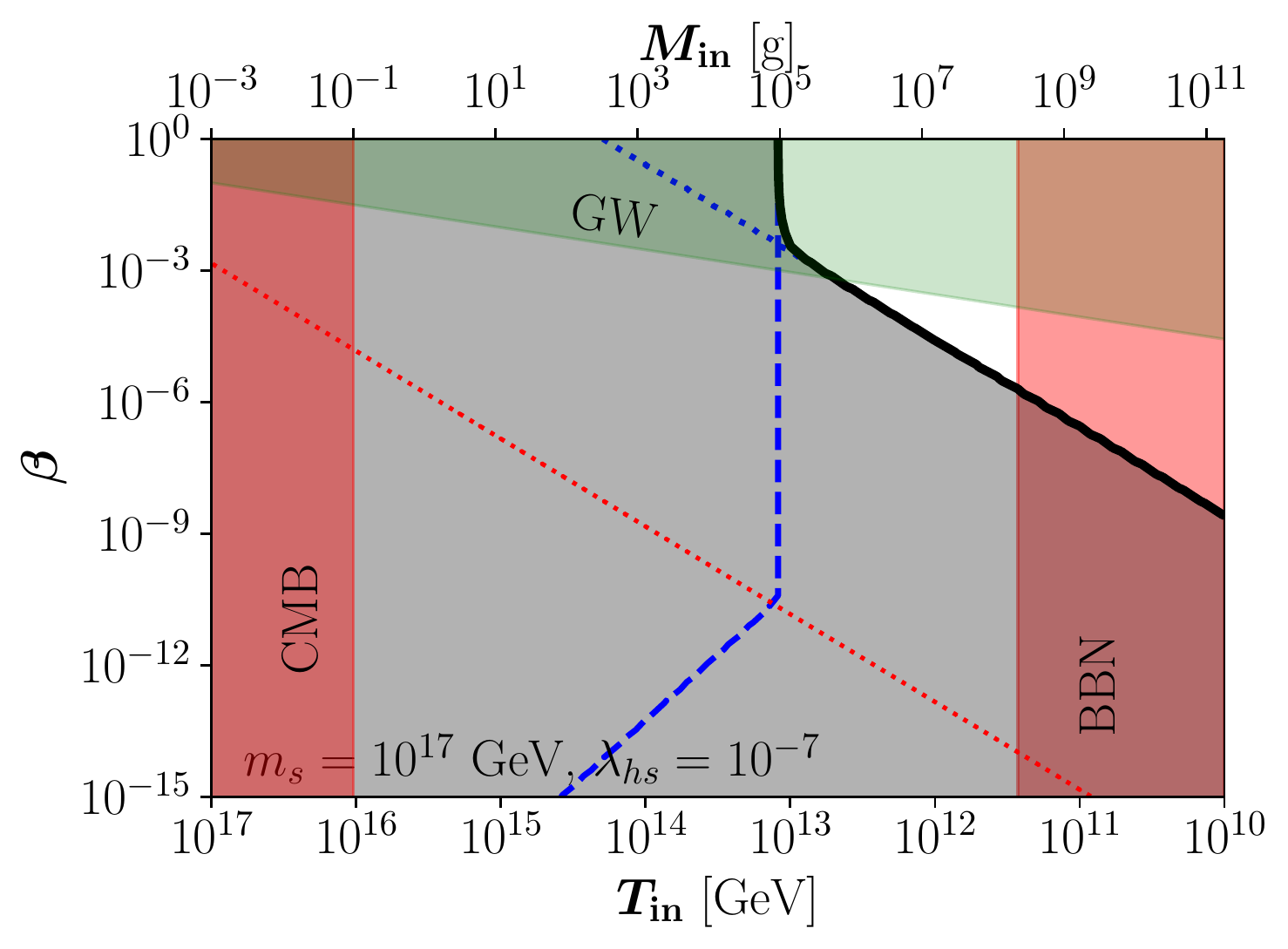}
	\includegraphics[scale=0.51]{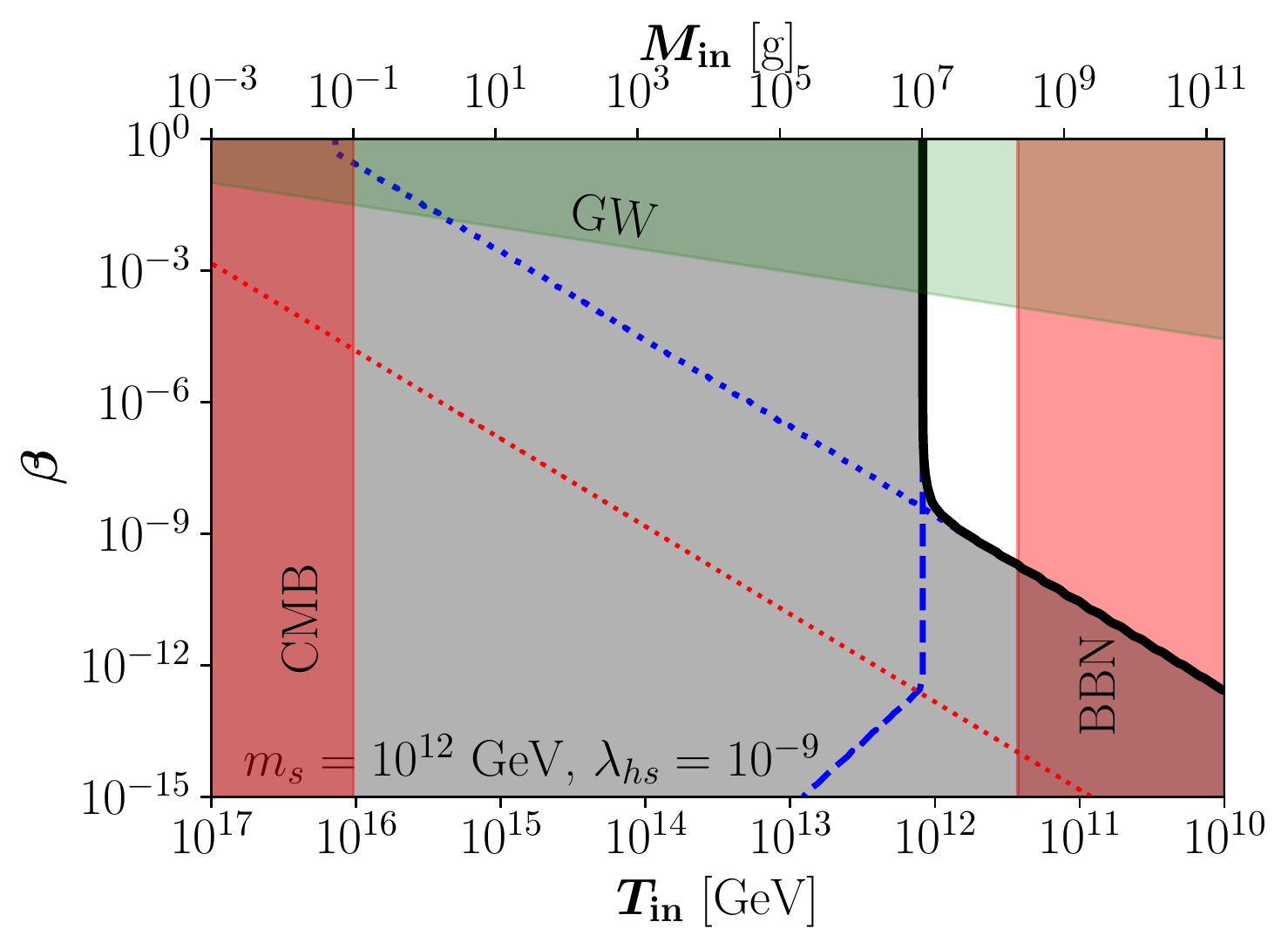}
	\includegraphics[scale=0.51]{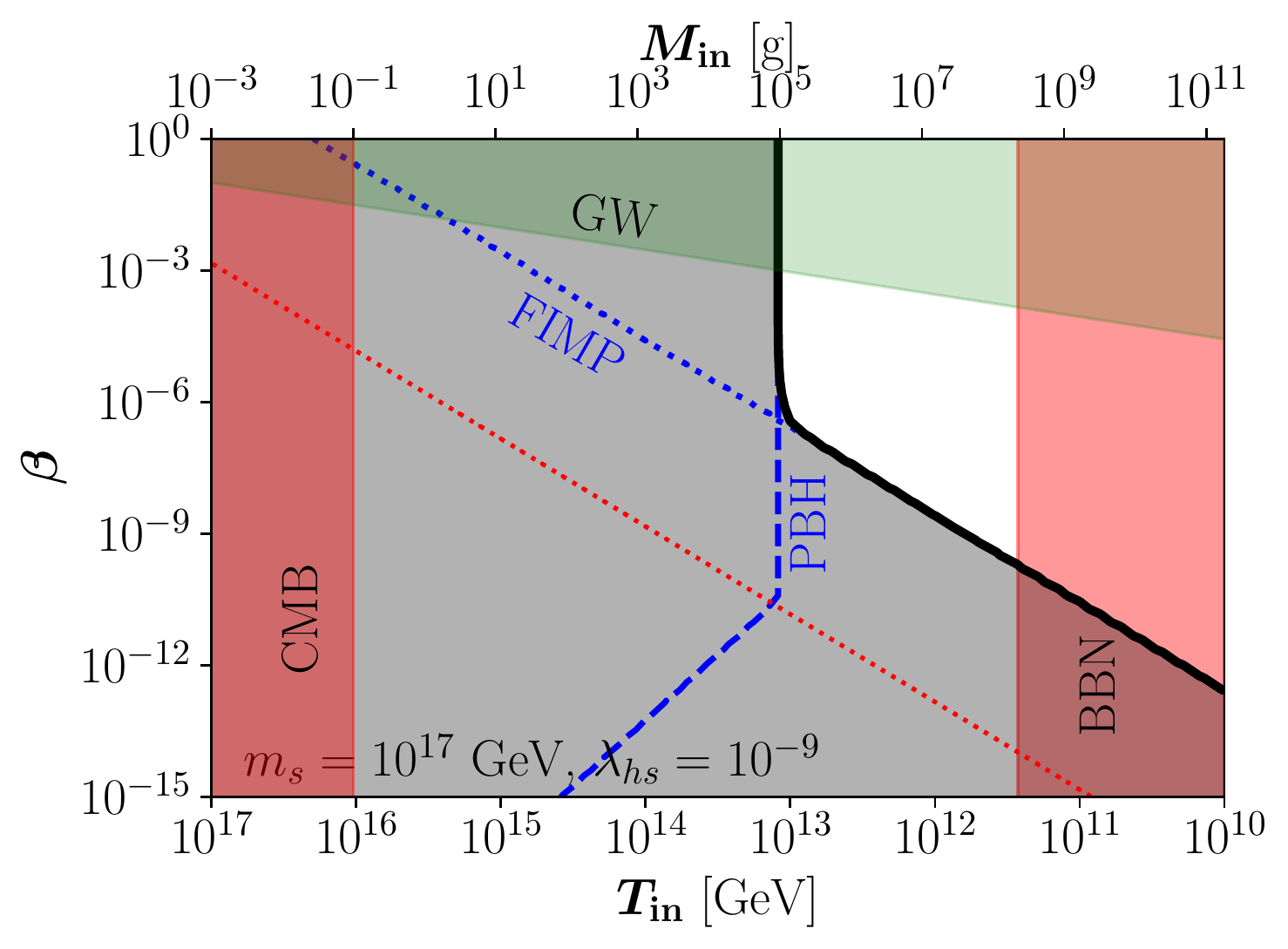}
	\caption{DM relic abundance due to the direct evaporation of PBHs (dashed blue lines) and the FIMP mechanism (dotted blue lines), for $\mdm = 10^{12}$~GeV (left panels) and $10^{17}$~GeV (right panels), and $\lhs = 10^{-7}$ (upper panels) and $10^{-9}$ (lower panels).
	The shaded areas are excluded by different observables described in the text}
	\label{fig:fimp2}
\end{figure}

Finally, there is another region, corresponding to $\mdm \gtrsim 10^9$~GeV, where DM can be produced in a PBH dominated era.
It is shown in Fig.~\ref{fig:fimp2}, for $\mdm = 10^{12}$~GeV (left panels) and $10^{17}$~GeV (right panels), and $\lhs = 10^{-7}$ (upper panels) and $10^{-9}$ (lower panels).
In this case, both the PBH and the FIMP production are viable, without being limited by the hot DM constraint.
We note that in this range of mass, the irreducible DM production via the gravitational UV freeze-in, where DM is generated via 2-to-2 annihilations of SM particles mediated by the exchange of massless gravitons in the $s$-channel, is very effective and can be dominant~\cite{Garny:2015sjg, Tang:2017hvq, Garny:2017kha, Bernal:2018qlk, Chianese:2020yjo, Bernal:2020ili}.
However, as this channel strongly depends on the details of the reheating dynamics, here it will be not considered.

\subsection{WIMP}
\begin{figure}
	\centering
	\includegraphics[scale=0.51]{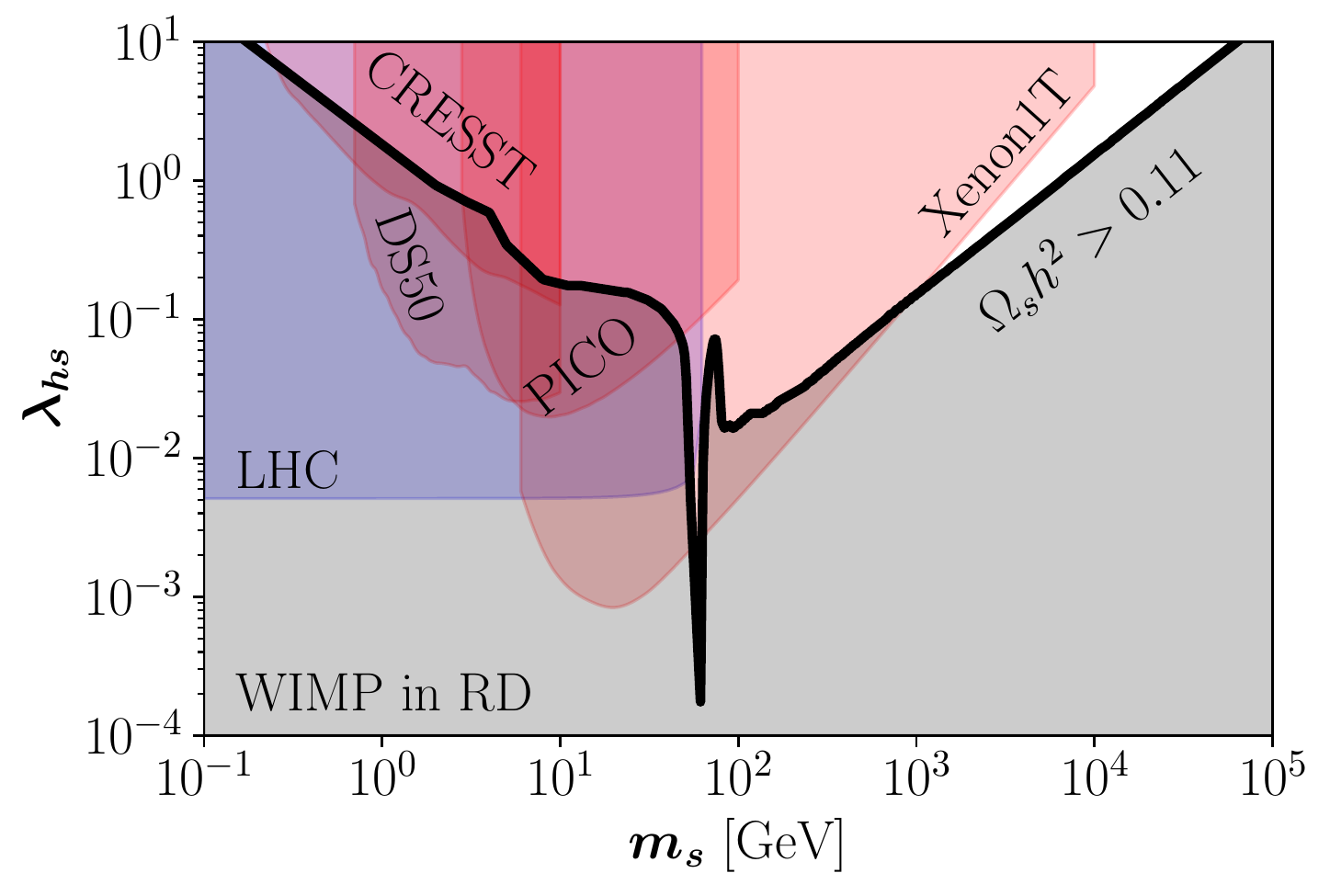}\\
	\includegraphics[scale=0.51]{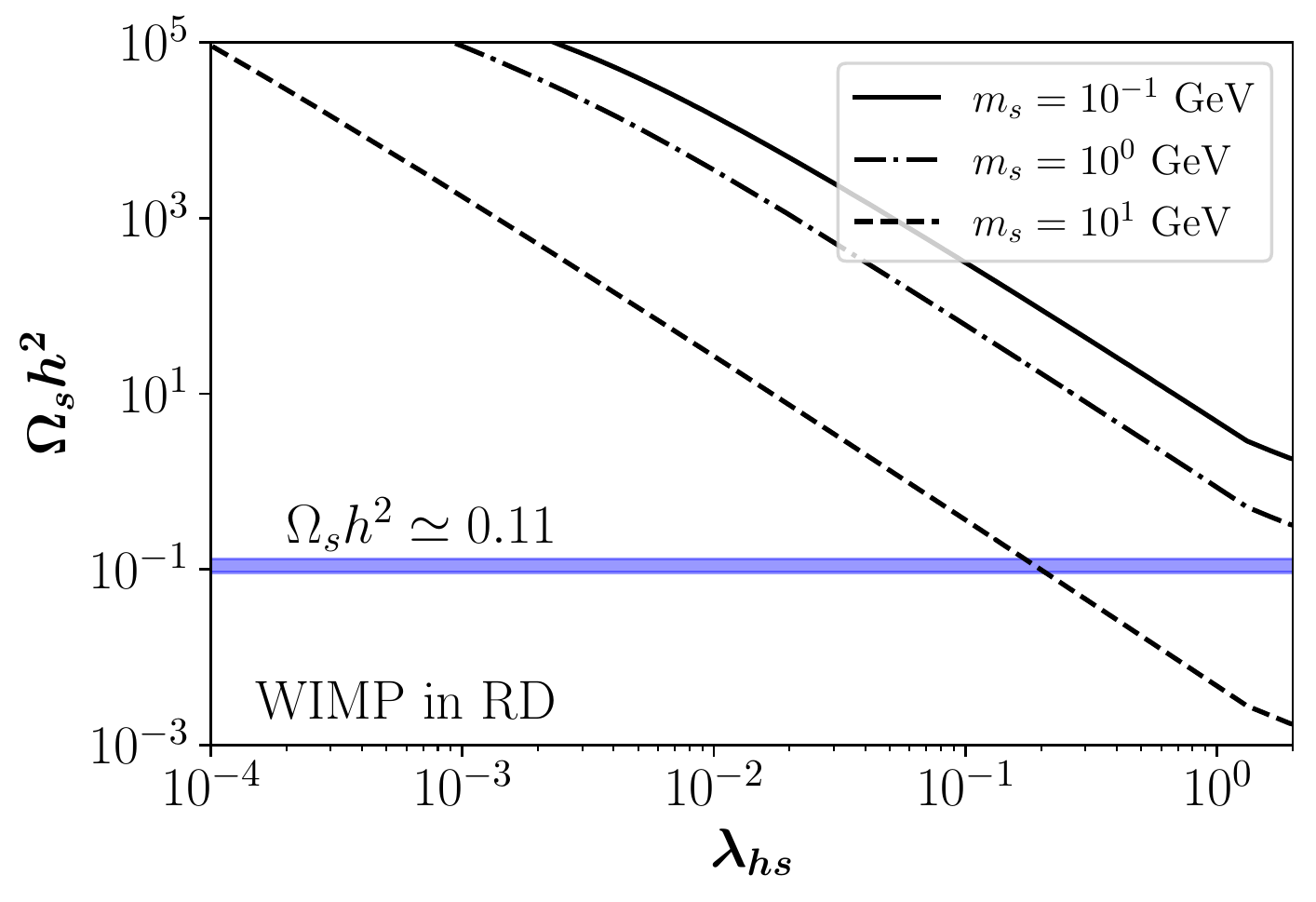}
	\includegraphics[scale=0.51]{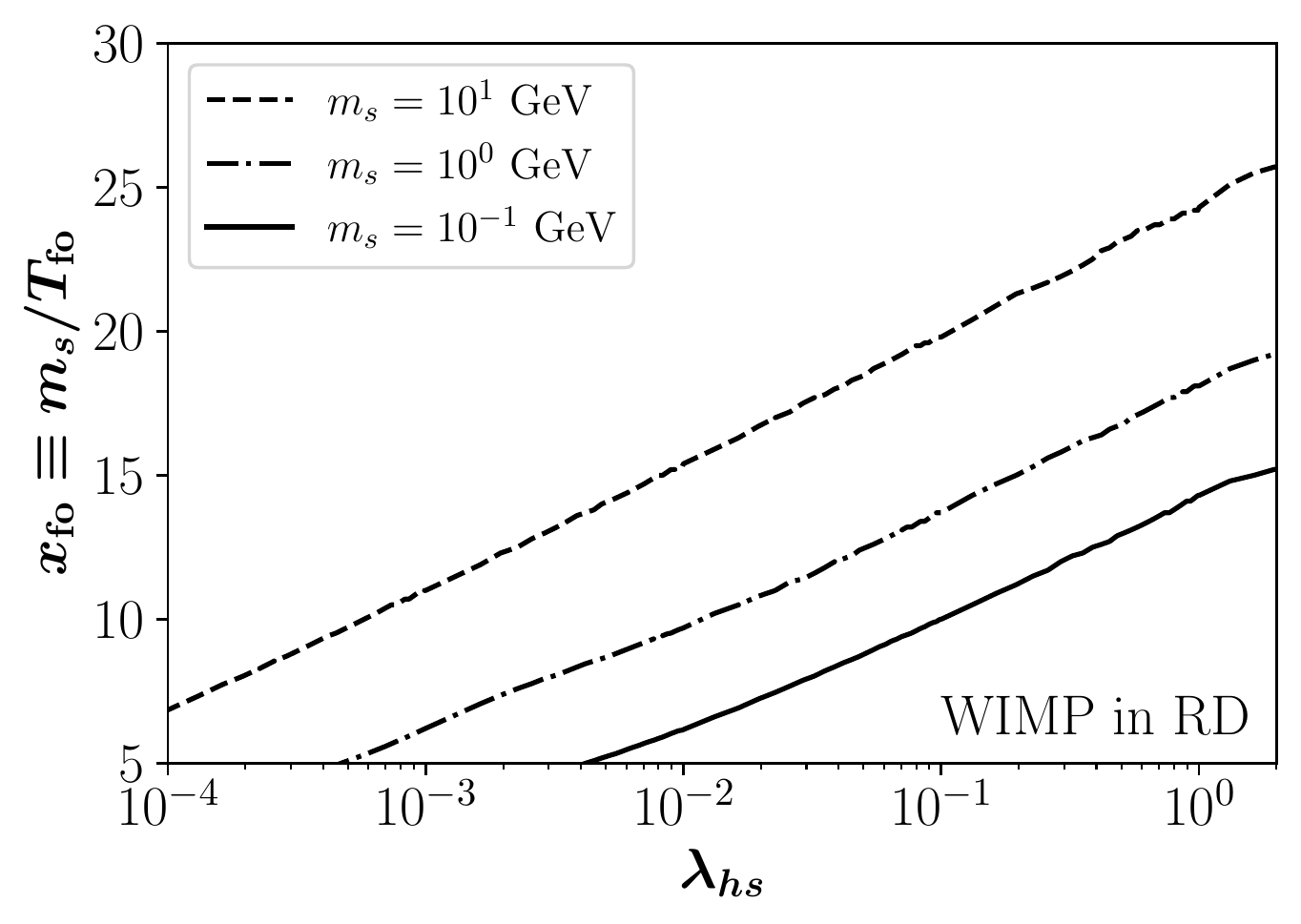}
	\caption{Upper panel: Portal coupling $\lhs$ required to reproduce the observed DM relic abundance as a function of the DM mass $\mdm$, in the WIMP scenario, assuming a standard cosmology (black line).
	Smaller couplings overclose the universe and are therefore excluded (gray area)
	Shaded red areas are in tension with direct detection experiments, and blue area with the invisible decay width of the Higgs as measured by ATLAS.
	The lower panels show the evolution of the DM abundance $\Omega_\text{DM}h^2$ (left) and the freeze-out parameter $x_\text{fo}\equiv \mdm/\Tfo$ (right) as a function of $\lhs$.
	}
	\label{fig:wimpRD}
\end{figure}

DM can also be generated in the early universe via the WIMP paradigm.
However, in the singlet scalar model, this mechanism is severely constrained.
The upper panel of Fig.~\ref{fig:wimpRD} shows with a thick black line the required values for the coupling $\lhs$ as a function of the DM mass, in the standard case of a universe dominated by SM radiation.
The lower panels depict the variation of the DM abundance $\Omega_sh^2$ (left panel) and the freeze-out temperature $T_\text{fo}$ (right panel), as a function of $\lhs$, for different DM masses $\mdm$.
In the figure, the gray-shaded area corresponding to larger couplings represents DM overproduction and is excluded.
Additionally, the red areas are in tension with different DM direct detection experiments: PICO-60~\cite{Amole:2019fdf}, CRESST-III~\cite{Abdelhameed:2019hmk}, DarkSide-50~\cite{Agnes:2018ves} and XENON1T~\cite{Aprile:2018dbl}.
Recent results from the ATLAS collaboration using 139~fb$^{-1}$ collisions at $\sqrt{s} = 13$~TeV bound the invisible Higgs boson branching ratio to be Br$_\text{inv}\leq 0.11$ at $95\%$~C.L.~\cite{ATLAS:2020kdi}.
The Higgs decay into a couple of DM particles contributes to its invisible decay and is given by
\begin{equation}
    \Gamma_{h\to ss} = \frac{\lhs^2\,v^2}{8\pi\,m_h}\sqrt{1-\frac{4\mdm^2}{m_h^2}}\,.
\end{equation}
Taking into account the total Higgs decay width $\Gamma_h\simeq 4.07$~MeV~\cite{Djouadi:2005gi}, it follows that $\lhs \lesssim 5\times10^{-3}$ for $\mdm \ll m_h/2$ (blue region).

In the WIMP scenario, the interplay between the PLANCK measurement of the DM relic density, the XENON1T upper bound on the elastic scattering cross section, and the invisible Higgs decay restricts the viable $\mdm$ range to be around the Higgs funnel region ($\mdm \simeq m_h/2$), or above $\mdm \gtrsim 1$~TeV until reaching the unitarity bound, at $\mdm \sim 100$~TeV~\cite{Griest:1989wd}. 

Similarly to the freeze-in case described in the previous section, PBH evaporation injects entropy to the SM bath, diluting the DM abundance.
However, contrary to the FIMP case, in the WIMP scenario smaller Higgs portal couplings are required if PBHs manage to dominate the expansion rate of the universe, compared to the RD case.

The WIMP scenario is considerably more involving.
If $\beta > \beta_c$, PBHs start to dominate the universe energy density at $T=\Teq$, and completely evaporate at $T=\Tev$.
As PBHs radiate SM particles, SM radiation is not expected to behave as free radiation, i.e., with $\rR \propto a^{-4}$. 
It is important to note that between $\Tc<T<\Tev$, PBHs dominate the evolution of SM radiation energy density: in that range SM radiation scales like $\rR \propto a^{-3/2}$, and therefore the SM temperature $T \propto a^{-3/8}$.
Having this in mind, one can differentiate four phenomenologically distinct regimes characterized by the temperature $\Tfo$ at which the DM freeze-out happens~\cite{Arias:2019uol}:
\begin{enumerate}
    \item[$i)$] $\Tfo \gg \Teq$: This case corresponds to the scenario where the DM freeze-out happens during radiation domination, and much earlier than the time when PBHs dominate the energy density of the universe.
    Here the final DM abundance is given by the standard production in a universe dominated by radiation, with a dilution due to the entropy injected by PBHs.
    \item[$ii)$] $\Teq \gg \Tfo \gg \Tc$: Alternatively, PBHs can dominate the Hubble expansion rate, but SM radiation still behaves like free radiation $\rR \propto a^{-4}$.
    \item[$iii)$] $\Tc \gg \Tfo \gg \Tev$: In this case, PBHs dominate both the Hubble expansion rate and the evolution of SM radiation: $\rR \propto a^{-3/2}$.
    \item[$iv)$] $\Tev \gg \Tfo$: Finally, even if PBHs give rise to a nonstandard cosmological era, as their evaporation happens at temperatures higher than the DM freeze-out, they have no effect on the WIMP dynamics.
    Moreover, all the DM radiated by PBHs thermalizes with the SM, and therefore does {\it not} contribute to the final DM relic abundance.
\end{enumerate}

A complete analysis of the WIMP production in a cosmology dominated by PBHs requires the numerical solution of a set of Boltzmann equations.
Here however, we follow a semianalytical approach, assuming an {\it instantaneous} PBH evaporation at $T=\Tev$.
In this case, regime $iii)$ can not be explored.
For the sake of completeness, in Appendix~\ref{sec:appendix} the relevant Boltzmann equations for the PBH energy density, the SM entropy density, and the DM number density are reported.

\begin{figure}
	\centering
	\includegraphics[scale=0.51]{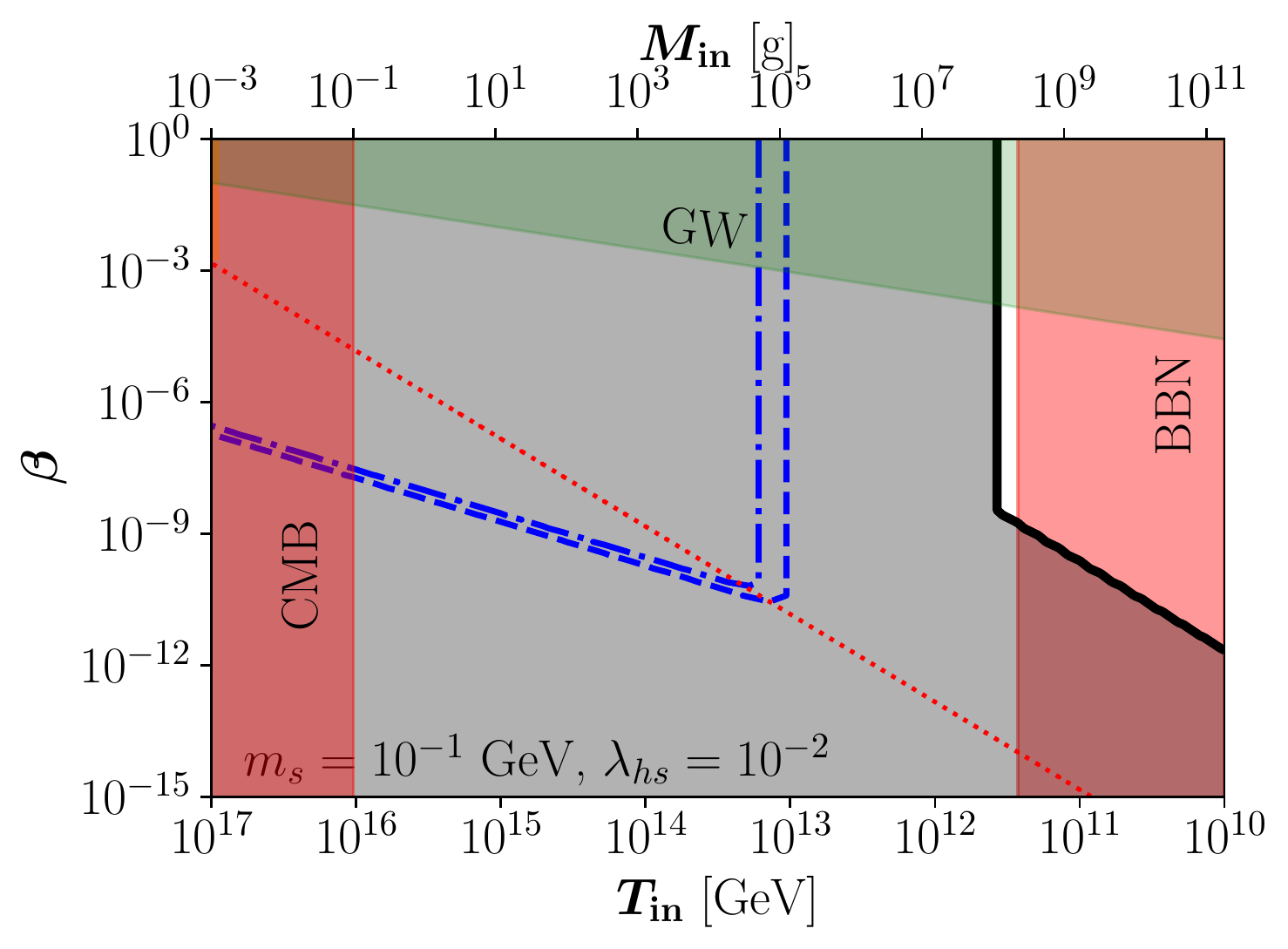}
	\includegraphics[scale=0.51]{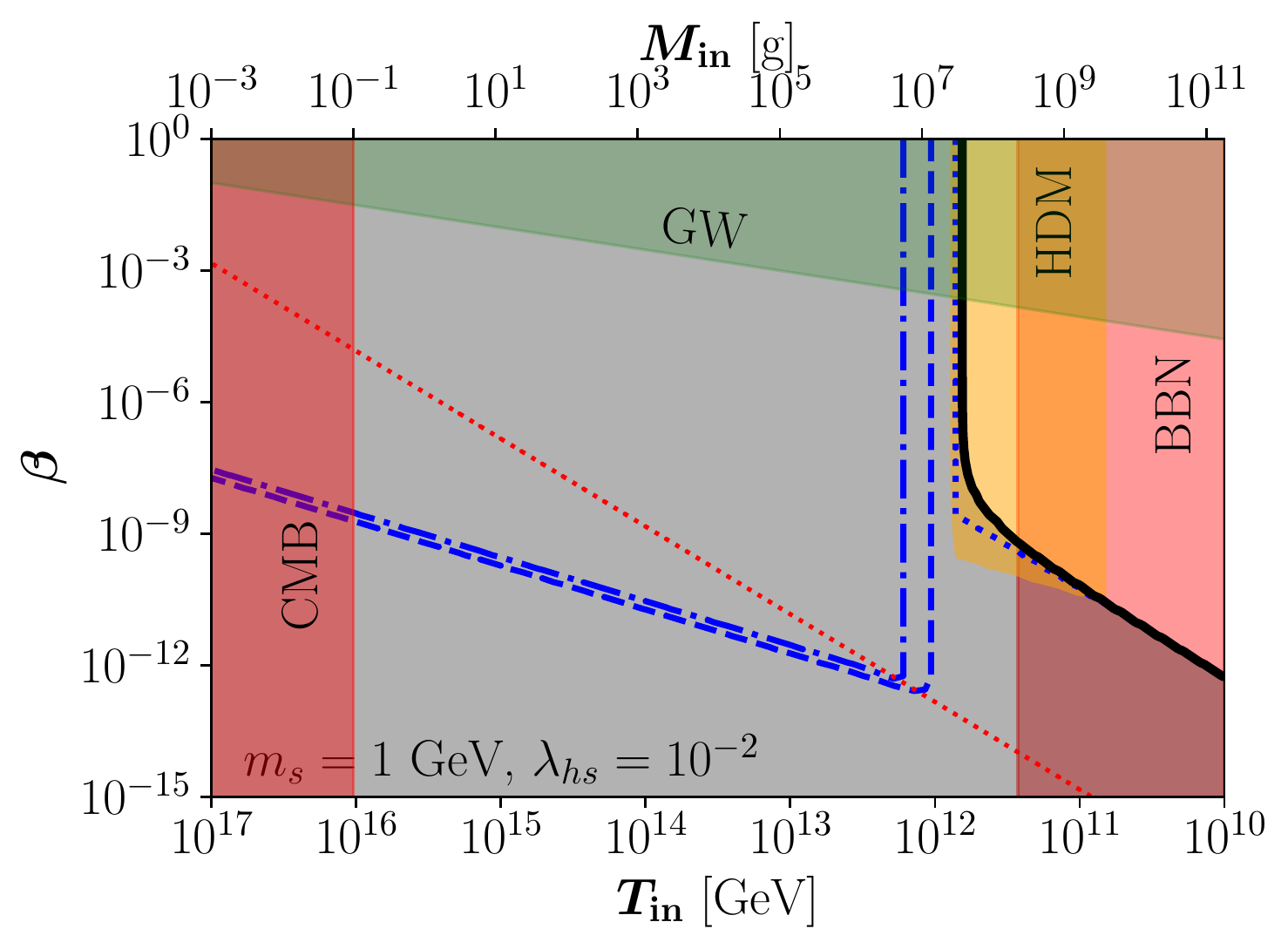}
	\includegraphics[scale=0.51]{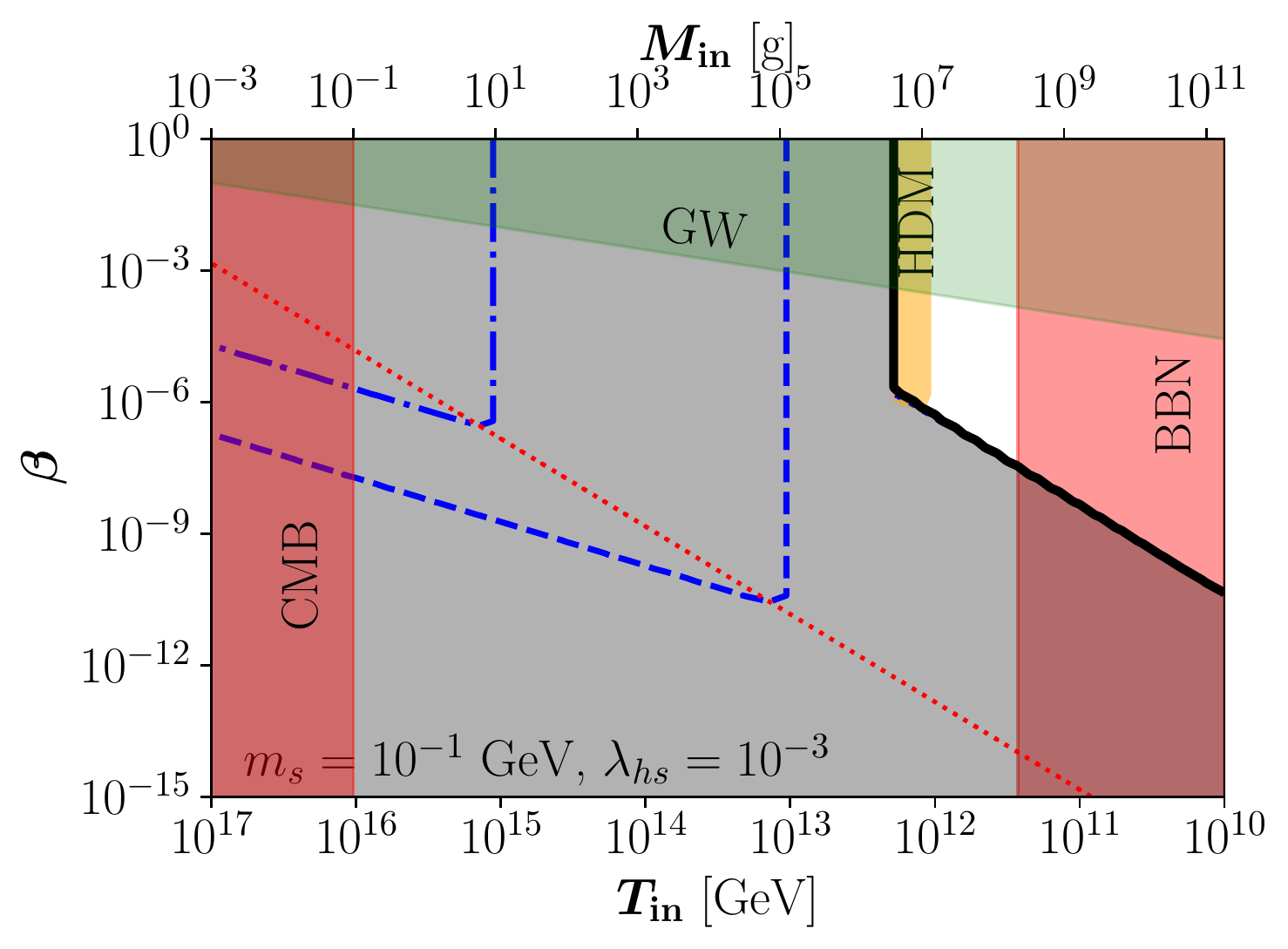}
	\includegraphics[scale=0.51]{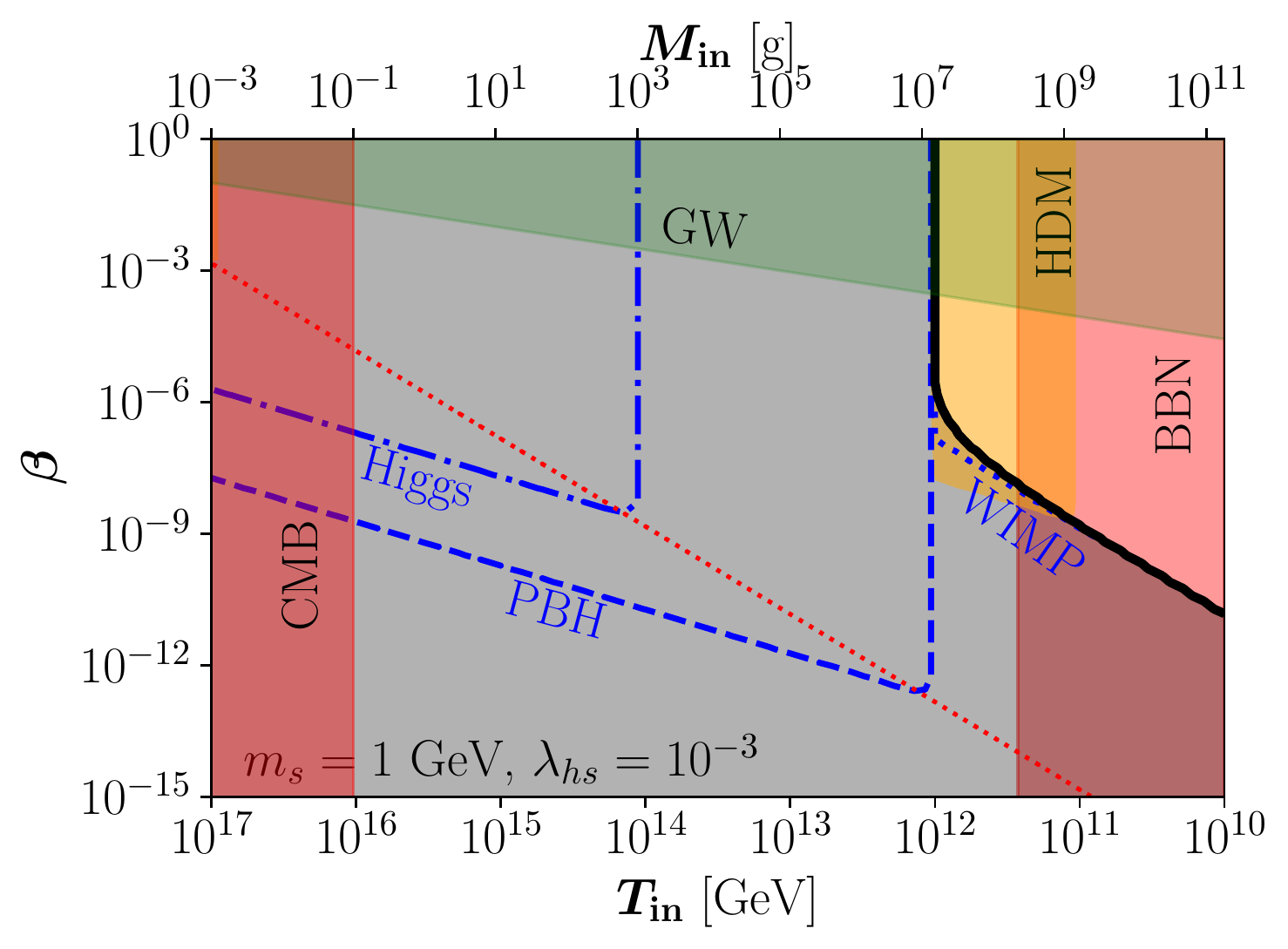}
	\caption{DM relic abundance due to the direct evaporation of PBHs (dashed blue lines), decay of Higgs bosons radiated by PBHs (dash-dotted blue lines), and the WIMP mechanism (dotted blue lines), for $\mdm = 10^{-1}$~GeV (left panels) and 1~GeV (right panels), and $\lhs = 10^{-2}$ (upper panels) and $10^{-3}$ (lower panels).
	}
	\label{fig:wimp}
\end{figure} 
In the case where PBHs never dominate the total energy density ($\beta < \beta_c$), the final DM abundance is given by the one of the WIMP mechanism, with an extra contribution coming from PBHs only if $\Tev < \Tfo$.
However, in the case opposite case where there is a PBH dominated era ($\beta > \beta_c$), entropy injection dilutes the produced DM, and therefore smaller Higgs portal couplings are typically required.
Figure~\ref{fig:wimp} shows the contributions to the DM relic abundance due to the direct evaporation of PBHs (dashed blue lines), the secondary production via decay of Higgs bosons radiated by PBHs (dash-dotted blue lines), and the WIMP mechanism (dotted blue lines), for $\mdm = 10^{-1}$~GeV (left panels) and 1~GeV (right panels), and $\lhs = 10^{-2}$ (upper panels) and $10^{-3}$ (lower panels).
The thick black lines correspond to the total contribution.
The values for the couplings were chosen so that they are smaller than the required values for a radiation-dominated universe, i.e., $\lhs \simeq 20$ for $\mdm =10^{-1}$~GeV, and $\lhs \simeq 2$ for $\mdm =1$~GeV.
We note that the secondary DM production via decay of Higgs bosons radiated by PBHs is now visible in the figure, however its contribution is typically subdominant.
Additionally, when PBHs generate a sizeable fraction of the total DM density, the hot DM constraint rules out large portions of the otherwise available parameter space.
We emphasize that the dilution allows the WIMP mechanism to be viable for lower Higgs portal couplings and for small masses in the MeV range, a parameter region that could be probed through invisible Higgs boson decays at the LHC in its high luminosity upgrade ($\lhs\gtrsim3\times10^{-3}$)~\cite{Cepeda:2019klc}, and the Future Circular Collider ($\lhs\gtrsim8\times10^{-4}$)~\cite{Benedikt:2018csr, deBlas:2019rxi}.  

\section{Conclusions} \label{sec:con}
The formation and subsequent evaporation of light primordial black holes (PBHs)
can lead to a departure from the standard cosmological history in the way they can dominate the evolution of the universe during a slot of time prior to the onset of the Big Bang nucleosynthesis. 
Under this premise and within the framework of the singlet scalar dark matter (DM) model, we have explored the impact of a BH domination at early times on the relic DM abundance when either the FIMP or WIMP production mechanism is operating along with the production via Hawking evaporation of the PBHs.
Thanks to the entropy delivered during PBH evaporation, large zones of the parameter space that were excluded for WIMPs and FIMPs, now become viable. In the case of the FIMP scenario, larger Higgs portal couplings than the required ones in a radiation-dominated universe are feasible (an increase of several orders of magnitude). 
On the other hand, when the singlet scalar is a WIMP relic, the regions featuring DM masses in the MeV range turn to be compatible.

The conclusions obtained in this work could be generalized to other FIMP and WIMP scenarios. For instance, the singlet fermionic DM model~\cite{Kim:2008pp} serves as a FIMP scenario~\cite{Klasen:2013ypa}  where a boost factor in the coupling controlling the DM relic density can be also expected. 
On the other hand, in the inert doublet model (IDM)~\cite{Deshpande:1977rw, Barbieri:2006dq, LopezHonorez:2006gr} when the DM candidate has a mass below $\sim40$~GeV the expected DM phenomenology resembles that of the singlet scalar DM model. Thus, one could expect that the DM masses around the MeV range become part of the viable DM mass regions of the IDM.   
All in all, larger (smaller) values than the required ones within a radiation domination era for the couplings entering in the DM relic density calculation  may be envisaged in FIMPs (WIMPs) scenarios. 

\section*{Acknowledgments}
The authors thank Xiaoyong Chu, Miguel Escudero and members of ``El Journal Club más sabroso'' for fruitful discussions.
NB received funding from Universidad Antonio Nariño grants 2018204, 2019101, and 2019248, the Spanish MINECO under grant FPA2017-84543-P, and the Patrimonio Autónomo - Fondo Nacional de Financiamiento para la Ciencia, la Tecnología y la Innovación Francisco José de Caldas (MinCiencias - Colombia) grant 80740-465-2020.
The work of OZ is supported by Sostenibilidad UdeA, the UdeA/CODI Grant 2017-16286, and by COLCIENCIAS through the Grant 1115-7765-7253.  
This project has received funding /support from the European Union's Horizon 2020 research and innovation programme under the Marie Skłodowska-Curie grant agreement No 860881-HIDDeN.

\appendix
\section{Boltzmann Equations} \label{sec:appendix}
For the sake of completeness, we report here the Boltzmann equations for the PBH energy density $\rbh$, the SM entropy density $s$, and the DM number density $\ndm$~\cite{Giudice:2000ex, Drees:2017iod, Masina:2020xhk}:
\begin{eqnarray}
    \frac{d\rbh}{dt}+3H\,\rbh&=&+\frac{\rbh}{\Mbh}\,\frac{d\Mbh}{dt}\,,\label{eq:BEPBH}\\
    \frac{ds}{dt} +3H\,s &=&-\frac{1}{T}\frac{\rbh}{\Mbh}\,\frac{d\Mbh}{dt} + 2\frac{E}{T} \sv \left[\ndm^2 - (\ndm^\text{eq})^2\right],\\
    \frac{d\ndm}{dt}+3H\,\ndm&=&+\frac{\rbh}{\Mbh}\,\frac{d\Ndm}{dt} - \sv \left[\ndm^2 - (\ndm^\text{eq})^2\right],\label{eq:BEDM0}
\end{eqnarray}
where $H^2=(\rR+\rbh+\rho_\text{DM})/(3M_P^2)$, $\ndm^\text{eq}$ is the DM number density at  equilibrium, $\sv$ is the 2-to-2 DM annihilation cross-section into SM particles, and $E \simeq \sqrt{\mdm^2 + 3 T^2}$ is the averaged energy per DM particle.
Additionally, the emission rate of DM particles per PBH is
\begin{equation}
    \frac{d\Ndm}{dt} = \int_0^\infty dE\, \frac{d^2 \Ndm}{dt\,dE}\,.
\end{equation}
As a final comment, we note that the source term in Eq.~\eqref{eq:BEPBH} can be rewritten as
\begin{equation}
    \frac{\rbh}{\Mbh}\,\frac{d\Mbh}{dt} = -\frac{\pi\,\gs(\Tbh)}{480}\frac{M_P^4}{\Mbh^3}\rbh = -\Gamma_\text{BH}(t)\,\rbh\,.
\end{equation}
Equation~\eqref{eq:BEPBH} is the same as the Boltzmann equation of a non-relativistic state with an effective {\it time-dependent} decay width $\Gamma_\text{BH}(t) = \frac{\pi\,\gs(\Tbh)}{480}\frac{M_P^4}{\Mbh^3(t)}$\,.

\bibliographystyle{JHEP}
\bibliography{biblio}

\end{document}